\documentclass[12pt]{iopart}
\usepackage{latexsym}

\begin{document}
\title{Resolving a gravitational wave memory paradox}

\author{David Garfinkle}
\address{Dept. of Physics, Oakland University,
Rochester, MI 48309, USA}
\address{and Michigan Center for Theoretical Physics, Randall Laboratory of Physics, University of Michigan, Ann Arbor, MI 48109-1120, USA}
\ead{garfinkl@oakland.edu}
\author{Istv\'an R\'acz}
\address{Wigner RCP, H-1121 Budapest, Konkoly Thege Mikl\'os \'ut 29-33, Hungary}
\ead{istvan@wigner.mta.hu}


\date{\today}

\begin{abstract}
Two different approaches to gravitational perturbation theory appear to give two different answers for the properties of gravitational wave memory.  We show that this contradiction is only apparent and the two approaches actually agree.

\end{abstract}


\maketitle

\section{Introduction}
\indent
Gravitational wave memory is a distortion in the gravitational wave detector that persists even after the wave has passed.  This effect was treated first in perturbation theory by Zel'dovich and Polnarev\cite{zeldovich}, and later in full nonlinear general relativity by Christodoulou.\cite{christodoulou}  The coordinate invariance of general relativity gives rise to a type of gauge invariance in perturbation theory. Here a gauge transformation is simply an infinitesimal coordinate transformation.  Thus a perturbed quantity transforms under a gauge transformation by addition of the Lie derivative of the background quantity in the direction of the coordinate transformation.  Therefore quantities that vanish in the background have perturbations that are gauge invariant.  In particular, in the usual weak field gravity perturbation theory, where the background is flat spacetime, the metric perturbation is not gauge invariant, but the Weyl tensor is gauge invariant.  When calculating a physical effect, such as gravitational wave memory, it is important that the results of the calculation are gauge invariant.  One way to insure this is to do the calculation using manifestly gauge invariant quantities.  Thus in \cite{lydiaandme} gravitational wave memory is treated using the perturbed Weyl tensor.  Another approach is to note how metric perturbations transform under gauge transformations and to find some combinations of metric perturbations that are gauge invariant.  This latter approach is the one most often used in treatments of gravitational waves in general, including gravitational wave memory.  A detailed exposition of this approach is given in \cite{scottandeanna}. 

However, as noted in\cite{istvan} the application of this second approach to gravitational wave memory gives rise to a paradox as follows: one of the gauge invariant quantities, $\Theta$, corresponds to the overall scale of the spatial metric.  If 
$\Theta$ is changed by a pulse of gravitational radiation, then there will be an overall isotropic distortion of the metric: an isotropic gravitational wave memory.  The treatment of \cite{istvan} gives a plausibility argument that one should expect $\Theta$ to be changed by a pulse of gravitational waves.  However, in the full nonlinear treatment of \cite{christodoulou} there is no isotropic memory, nor is there any isotropic memory in the perturbative treatment of \cite{lydiaandme}.

In this paper we will resolve this paradox.  Section 2 presents the gauge invariant metric variables and their relation to the components of stress-energy and then summarizes the plausibility argument of \cite{istvan} in favor of isotropic memory.  In section 3 we calculate the change in $\Theta$ due to the passage of a gravitational wave and show that it vanishes.  Our conclusions are given in section 4.

\section{Gauge Invariant Metric Quantities}         

To first order in perturbation theory the metric is ${g_{\alpha \beta}} = {\eta _{\alpha \beta}} + 
{h_{\alpha \beta}}$ where $\eta _{\alpha \beta}$ is the background flat metric and $h_{\alpha \beta}$ is small.  We use greek letters for spacetime indices and latin letters for spatial indices. 
 Under a gauge transformation determined by a vector $\xi ^\alpha$, the metric perturbation transforms as ${h_{\alpha \beta}} \to {h_{\alpha \beta}} + 2 {\partial _{(\alpha}}{\xi_{\beta )}}$. Thus the metric components are certainly not gauge invariant.  However, gauge invariant quantities can be constructed from the metric.  We begin by presenting this standard construction, recapitulating the presentation of  \cite{scottandeanna}.  First the metric perturbation $h_{\alpha \beta}$ is decomposed into scalars, divergence free spatial vectors, and a divergence free and trace free spatial tensor as follows:
\begin{eqnarray}
{h_{tt}} = 2 \phi
\label{hscalar}
\\
{h_{ti}} = {\beta _i} + {\partial _i} \gamma
\label{hvector}
\\
{h_{ij}} = {h^{TT} _{ij}} + {\textstyle {\frac 1 3}} H {\delta _{ij}} + {\partial _{(i}}{\varepsilon _{j)}} + 
\left ( {\partial _i}{\partial _j} - {\textstyle {\frac 1 3}} {\delta _{ij}} {\nabla ^2} \right ) \lambda
\label{htensor}
\end{eqnarray}
Given $h_{\alpha \beta}$ the decomposition is done as follows: the divergence of eqn. (\ref{hvector}) along with the condition that ${\partial ^i}{\beta _i}=0$ provides an elliptic equation for $\gamma$ which determines $\gamma$ subject to the boundary condition that $\gamma $ vanishes at infinity.  Similarly, two divergences of eqn. (\ref{htensor}) provides an elliptic equation that determines $\lambda$, and then one divergence of eqn. (\ref{htensor}) provides an elliptic equation that determines $\varepsilon _i$.  

 The quantity $h^{TT}_{ij}$ is gauge invariant.  The other quantities in the decomposition are not gauge invariant, but particular combinations of them are gauge invariant.  Define the quantities $\Phi, \, \Theta$ and $\Xi _i$ by
\begin{eqnarray}
\Phi \equiv  - \phi + {\partial _t} \gamma - {\textstyle {\frac 1 2}} {\partial ^2 _t}\lambda
\\
\Theta \equiv {\textstyle {\frac 1 3}} (H - {\nabla ^2} \lambda)
\\
{\Xi _i} \equiv {\beta _i} - {\textstyle {\frac 1 2}} {\partial _t} {\varepsilon _i}
\end{eqnarray}
Then $\Phi, \, \Theta$ and $\Xi _i$ are gauge invariant.

To see how the gauge invariant metric perturbations are determined by the stress-energy, one begins by decomposing the stress-energy into scalars, divergence free vectors and a divergence free, trace free tensor as follows:
\begin{eqnarray}
{T_{tt}} = \rho
\\
{T_{ti}} = {S_i} + {\partial _i} S 
\\
{T_{ij}} = {\sigma _{ij}} + P {\delta _{ij}} + {\partial _{(i}}{\sigma _{j)}} + \left ( {\partial _i}{\partial _j} - {\textstyle {\frac 1 3}} {\delta _{ij}} {\nabla ^2} \right ) \sigma
\end{eqnarray}
The decomposition is performed in the same way as for the metric perturbation.  However, since the background has vanishing stress-energy, it follows that the stress-energy is gauge invariant, and therefore the quantities into which it is decomposed are also gauge invariant.  The linearized Einstein field equations yield the following:
\begin{eqnarray}
{\nabla ^2} \Theta = - 8 \pi \rho
\label{PTheta}
\\
{\nabla ^2} \Phi = 4 \pi (\rho + 3 P - 3 {\partial _t} S)
\label{PPhi}
\\
{\nabla ^2} {\Xi _i} = - 16 \pi {S_i}
\label{PXi}
\\
\Box {h^{TT} _{ij}} = - 16 \pi {\sigma _{ij}}
\label{waveh}
\end{eqnarray}
Note that although the variables in eqns. (\ref{PTheta}-\ref{waveh}) are gauge invariant, they are also non-local.  This is clear for $\Theta, \, \Phi$ and $\Xi _i$ since they are determined  by Poisson equations.  However, even $h^{TT}_{ij}$, which is determined by a wave equation, is non-local.  This is because the quantity $\sigma _{ij}$, which is the source for $h^{TT}_{ij}$, has pieces which are determined by Poisson equations and therefore the source is non-local.

We now turn to the argument of \cite{istvan} for isotropic memory.  The mirrors of the gravitational wave interferometer follow geodesics, so their separation $D_i$ obeys the geodesic deviation equation
\begin{equation}
{\frac {{d^2}{D_i}} {d{t^2}}} = - {R_{titj}}{D^j}
\label{geodev}
\end{equation}
Thus, to find the gravitational wave memory, one first computes the leading $1/r$ piece of the $titj$ component of the Riemann tensor and then integrates twice with respect to time.  As shown in \cite{scottandeanna} this component of the Riemann tensor is given by
\begin{equation}
{R_{titj}} = - {\textstyle {\frac 1 2}} {\partial ^2 _t} \left  ( {h^{TT} _{ij}} +
\Theta {\delta _{ij}} \right ) + {\partial _i}{\partial _j} \Phi + {\partial _t} {\partial _{(i}}{\Xi_{j)}}
\label{Riemann}
\end{equation}
From eqns. (\ref{PTheta}-\ref{waveh}) it follows that ${h^{TT}_{ij}}, \, \Theta, \, \Phi$ and $\Xi_i$ all have a $1/r$ piece.  However, the presence of spatial derivatives in the last two terms of eqn. (\ref{Riemann})
makes these terms higher order in $1/r$.  Thus to lowest order in $1/r$ we have
\begin{equation}
{R_{titj}} = - {\textstyle {\frac 1 2}} {\partial ^2 _t} \left  ( {h^{TT} _{ij}} +  
\Theta {\delta _{ij}} \right ) 
\label{Riemann2}
\end{equation}
Then using eqn. (\ref{Riemann2}) in eqn. (\ref{geodev}) and integrating twice with respect to time we find that the change in separation is given by
\begin{equation}
\Delta {D_i} = {\textstyle {\frac 1 2}} {D^j} \Delta {h^{TT}_{ij}} + {\textstyle {\frac 1 2}} {D_i} \Delta \Theta
\label{memory}
\end{equation}
Here for any quantity $f$ the quantity $\Delta f$ is the change in $f$ between the begining and end of the gravitational wave train being detected.   Eqn. (\ref{memory}) differs from the usual expression for gravitational wave memory by the presence of the second term: the isotropic memory.  Thus either the usual expression for gravitational wave memory is wrong or there is some reason why the isotropic memory vanishes. If the observer were at a distance large compared to the distance of the stress-energy, then it would follow from equation (\ref{PTheta}) that $\Theta =2M/r$ to leading order in $1/r$ where $M$ is the total mass of the spacetime.  Since mass is conserved, it would then follow that  
the isotropic memory vanishes.  And  indeed, in \cite{scottandeanna} it is claimed that isotropic memory is ruled out because $\Theta =2M/r$.  However, it is easy to see that the observer is not at a distance large compared to the stress-energy.  This is certainly true for the effective stress-energy of the gravitational waves themselves, since they travel outward at the speed of light.  It is also true for other types of radiation emitted by the same process that makes the gravitational radiation, {\it e.g.} the electromagnetic radiation and neutrinos produced in core collapse supernovae.  Furthermore, the emitted radiation will lead to a recoil of the objects that make up the source stress-energy, and even these recoiling objects are not confined in a finite region: in a time $t$ an object recoiling with a speed $v_A$ travels a distance of ${v_A}t$ and therefore reaches a large distance at a late time.  In particular, \cite{istvan} considers a pulse of radiation that travels at the speed of light and therefore crosses the observer's position at the same time as the signal that makes the gravitational wave memory.  Since the stress-energy of the pulse is inside the observer's radius before the gravitational wave train passes and is outside the observer's radius after the wave train passes, one should expect a change in $\Theta$ coinciding with the time of the observation of the gravitational wave.  In other words, one should expect an isotropic component of gravitational wave memory.

\section{Isotropic Memory?}

Though the argument of \cite{istvan} is plausible, there are other approaches to gravitational wave memory that do not give an isotropic component of gravitational wave memory.  In particular, the treatment of \cite{christodoulou}, using the full nonlinear vacuum Einstein equations does not have an isotropic component of memory, nor do the generalizations of the result of \cite{christodoulou} to the case of the Einstein-Maxwell equations \cite{lydiaem} or the Einstein equations with neutrinos modeled as a null fluid.\cite{lydianull}  Futhermore a perturbative treatment of memory using the Weyl tensor rather than metric perturbations\cite{lydiaandme} also does not have an isotropic component of memory.  Thus we are left with a paradox: a perturbative and gauge invariant method based on metric variables seems to give an isotropic component of gravitational wave memory, while neither the full nonlinear Einstein equations nor a perturbative method using the Weyl tensor have this result.

In order to resolve this paradox, we will calculate the quantity $\Theta$ to see whether there is isotropic memory.  Using the Green's function for eqn. (\ref{PTheta}) we have
\begin{equation}
\Theta (t,{\vec r}) = 2 \int {d^3} {r '} {{\rho (t,{{\vec r}\, '})} \over {|{\vec r} -{ {\vec r}\, '}|}} \; .
\label{ThetaGreen}
\end{equation}
We consider a finite gravitational wave train produced at a finite time and then detected by an observer at large distance.  Because the wave travels at the speed of light, it follows that the wave is detected at a late time.  Since $\Theta$ satisfies a Poisson equation, it follows that in order to find $\Theta$ at late times, we only need the behavior of $\rho$ at late times.  As in\cite{lydiaandme} we will assume that at late times the stress-energy consists of a pulse of outgoing radiation plus a set of recoiling objects widely separated from each other and each traveling at constant velocity.   We can then write $\Theta = {\Theta_1}+{\Theta_2}$ where 
the recoiling objects are the source for $\Theta_1$ and the pulse of outgoing radiation is the source for $\Theta_2$.

We begin by calculating $\Theta_1$.  First consider one of the recoling objects with energy $E_A$ and velocity ${\vec v}_A$.  Define the retarded time $u=t-r$.  Then since the gravitational wave train has finite $u$ but is detected at large $r$, it follows that $t=r+O(1)$ and therefore that the position of the object is ${{\vec r}_A} =  r {{\vec v}_A} + O(1)$.  It then follows from eqn. (\ref{ThetaGreen}) that to leading order in $1/r$
\begin{equation}
{\Theta_1} = {\frac 2 r} {\sum _A} {\frac {E_A}  {\sqrt {1 + {v_A ^2} - 2 {\hat r}\cdot {{\vec v}_A}}}}
\label{Theta1result}
\end{equation}

We now calculate $\Theta_2$.
Let $\Omega$ be the point on the unit two-sphere corresponding to the direction of $\vec r$.  Then $\vec r$ is determined by $r$ and $\Omega$.  Also introduce the corresponding notation for
${\vec r} \, '$.   Then eqn. (\ref{ThetaGreen}) becomes
\begin{equation}
\Theta (t,r,\Omega) = 2 \int d {\Omega '} \, {\int _0 ^\infty} {{({r'})}^2} d {r '} 
 {{\rho (t,{r'},{\Omega '})} \over {|{\vec r} -{ {\vec r} \, '}|}} \; .
\label{ThetaGreen2}
\end{equation}
For outgoing radiation at late time, as shown in \cite{lydiaandme} the leading order behavior of $\rho$ is
\begin{equation}
\rho (t,{r'},{\Omega '}) = {{(r ' )}^{-2}} L(t- {r '},{\Omega '}) \; ,
\label{rhorad}
\end{equation}
where $L$ is the power radiated per unit solid angle. Here the first argument of $L$ is the time at which the radiation is emitted.  Since this time is $O(1)$ and since $t=r+O(1)$ it follows that $\rho$ vanishes unless $r={r'}+O(1)$.  Therefore to lowest order in $1/r$ we have
\begin{equation}
{1 \over {|{\vec r} -{ {\vec r}\, '}|}} = {1 \over {2 r \sin (\psi /2)}}
\label{rdiff}
\end{equation}
where $\psi$ is the angle between $\vec r$ and ${\vec r}\, '$.  
Then using eqns. (\ref{rhorad}) and (\ref{rdiff}) in eqn. (\ref{ThetaGreen2}) we obtain
\begin{equation}
{\Theta_2} (t,r,\Omega) = {1 \over r} \int d {\Omega '} \, {\int _0 ^\infty}  d {r '} 
 {{L(t- {r '},{\Omega '}) } \over {\sin (\psi /2)}} \; .
\label{Theta2prelim}
\end{equation}
Now define the quantity $F(\Omega)$ by
\begin{equation}
F (\Omega) = {\int _{-\infty } ^\infty} d u \, L(u,\Omega)
\label{Fdef}
\end{equation}
That is, since $L$ is the power radiated per unit solid angle, it follows that $F$ is the energy radiated per unit solid angle.  Then using eqn. (\ref{Fdef}) in eqn. (\ref{Theta2prelim}) we obtain
\begin{equation}
{\Theta_2} (t,r,\Omega) = {1 \over r} \int d {\Omega '} {{F({\Omega '})}\over {\sin (\psi/2)}}
\label{Theta2result}
\end{equation}
It is clear from eqns. (\ref{Theta1result}) and (\ref{Theta2result}) that $\Theta$ is not simply equal to $2M/r$.  However, it is also clear from these equations that (to leading order in $1/r$) $\Theta$ has no dependence on time.  It therefore follows that $\Delta \Theta = 0$ and therefore there is no isotropic component of gravitational wave memory.  The paradox is resolved.

\section{Conclusions}

Though the calculation of the previous section resolves the paradox of isotropic gravitational wave memory, it is also helpful to explain what went wrong with the plausibility argument of \cite{istvan}.   Since $\Theta$ is not simply $2M/r$ there is no conservation law to enforce the constancy in time of $\Theta$.  So why is it constant?  First note that since the observation of gravitational waves takes place at late times and since $\Theta$ is determined by a Poisson equation, it follows that $\Theta$ is determined by the late time behavior of the source.  But at late times the source consists of a set of objects in steady straight line motion and a bubble of radiation expanding at a constant rate.  So perhaps it is not so strange that this steady late time state of the source leads to a constant $\Theta$.  The most surprising aspect of the constancy of $\Theta$ is that the time of the observation includes the time when the bubble of radiation crosses the position of the observer, and yet this still does not lead to a change in $\Theta$.  However, here the surprise is due to our intuition relying on the example of the Newtonian gravitational field of a thin uniform shell: the gravitational field has a magnitude of $M/{r^2}$ when one is outside the shell and zero when one is inside the shell.  Thus we might expect some similar change in $\Theta$ as the shell of radiation crosses the observer's position.  Here, however it is helpul to note that $\Theta$ is analogous to the gravitational potential, not the gravitational field.  Even in the Newtonian case there is no change in the potential when the shell is crossed.  (For an expanding shell, the potential inside the shell changes as the shell expands; however this effect is higher order in $1/r$ and can therefore be neglected for our purposes).

Finally we note that this tricky conceptual issue of (the absence of) isotropic gravitational wave memory is due to an approach that uses metric perturbations to describe gravitational wave memory.  The  use of metric perturbations requires that one give up either gauge invariance or locality of the quantities used to describe the radiation.  In contrast, one can describe both the propagation of gravitational waves and their effects on gravitational wave detectors using Weyl curvature components as the basic quantities.  These quantities are both gauge invariant and local.  Though the use of metric perturbations is traditional, and in some cases it may be more convenient for calculations, nonetheless we believe that conceptually there is a great advantage to describing gravitational waves using curvature.

\ack
DG was supported by NSF grant number PHY-1205202 to Oakland University.  IR was supported by the European Union and the State of Hungary, co-financed by the European Social Fund in the framework of T\'AMOP-4.2.4.A/2-11/1-2012-0001 ``National Excellence Program.''  This research was supported in part by Perimeter Institute for Theoretical Physics.  Research at Perimeter Institute is supported by the government of Canada through Industry Canada and by the Province of Ontario through the Ministry of Economic Development \& Innovation.

\end{document}